\documentclass[showpacs,aps,twocolumn]{revtex4}
\usepackage{bm}
\usepackage{amsmath}
\usepackage{graphicx}
\usepackage{subfigure}
\usepackage[usenames,dvipsnames]{color}
\definecolor{darkblue}{RGB}{0,0,196}
\usepackage[colorlinks=true,linkcolor=darkblue,citecolor=darkblue,urlcolor=darkblue]{hyperref}
\usepackage{setspace}
\usepackage{hyperref}
\usepackage{xcolor}
\hypersetup{
  colorlinks   = true, 
  urlcolor     = red, 
  linkcolor    = blue, 
  citecolor   = blue 
}
\usepackage{footmisc}
\usepackage[makeroom]{cancel}
\usepackage{comment}
\usepackage{lineno}
\def\be{\begin{equation}}
\def\ee{\end{equation}}
\def\ba{\begin{eqnarray}}
\def\ea{\end{eqnarray}}
\usepackage{graphicx}
\usepackage{amsmath,bbm}
\usepackage{amssymb,bm}
\usepackage{yfonts}

\begin{document}
\title{First implementation of transverse spherocity analysis for heavy-ion collisions at the Large Hadron Collider energies}
\author{Neelkamal Mallick}
\author{Sushanta Tripathy}
\author{Raghunath Sahoo\footnote{Corresponding author: $Raghunath.Sahoo@cern.ch$}}
\affiliation{Discipline of Physics, School of Basic Sciences, Indian Institute of Technology Indore, Simrol, Indore 453552, India}
\author{Antonio Ortiz}
\affiliation{Instituto de Ciencias Nucleares, UNAM, Deleg. Coyoac\'{a}n, Ciudad de M\'{e}xico 04510}

\begin{abstract}
\noindent
Transverse spherocity, an event shape observable, has a very unique capability to separate the events based on their geometrical shape, {\it i.e.} jetty and isotropic. In this work, we use transverse spherocity for the first time in heavy-ion collisions using A Multi-Phase Transport Model (AMPT). We obtain the transverse momentum spectra, integrated yield, mean transverse momentum and azimuthal anisotropy for identified particles in Xe-Xe collisions at $\sqrt{s_{\rm{NN}}} = 5.44$~TeV and Pb-Pb collisions at $\sqrt{s_{\rm{NN}}} = 5.02$~TeV. The indication of collectivity in heavy-ion collisions can be clearly seen while comparing the transverse momentum spectra from jetty and isotropic events. The elliptic flow as a function of transverse spherocity shows that the isotropic events have nearly zero elliptic flow and the elliptic flow is mostly dominated by the jetty events. This study will pave a way to focus on jetty events in heavy-ion collisions in order to investigate jet medium modification and jet hadro-chemistry in a sophisticated manner.

 
\end{abstract}
\date{\today}
\maketitle 

\section{Introduction}
\label{intro}
A deconfined state of quarks and gluons, also known as Quark Gluon Plasma (QGP), is believed to be produced in ultra-relativistic heavy-ion collisions at the Large Hadron Collider (LHC) at CERN, Switzerland and Relativistic heavy-ion collider (RHIC) at BNL, USA. However, we do not have any direct evidence of possible QGP formation due to its very short lifetime instead several indirect signatures such as strangeness enhancement, direct photon measurements, elliptic flow etc. suggest that formation of QGP is highly probable in such collisions. Traditionally, the results from collisions of protons at RHIC and the LHC are considered as a baseline for the results obtained for heavy-ion collisions. Recent measurements in $pp$ collisions from LHC such as strangeness enhancement~\cite{ALICE:2017jyt}, ridge-like structures~\cite{Khachatryan:2016txc} have surprised the scientific community. The results were surprising because with merely 20-30 final state charged particle multiplicity at mid-rapidity, it is very hard to believe of a possible equilibrated medium formation in such collisions. There are several observations regarding a threshold in event multiplicity ($\frac{dN_{\rm ch}}{d\eta}\big|_{|\eta| < 1} \simeq 20$) for the formation of a system with different behavior~\cite{Thakur:2017kpv,Sahu:2019tch,Sahu:2020nbu,Sharma:2018jqf,Campanini:2011bj}. A possible QGP-droplet in small collision systems would have serious concerns about the results from heavy-ion collisions which uses $pp$ collisions as baseline. To understand the dynamics of small collision systems, an event shape observable, transverse spherocity, has been introduced recently~\cite{Cuautle:2014yda,Cuautle:2015kra,Salam:2009jx,Bencedi:2018ctm,Banfi:2010xy}. From these studies, it was observed that transverse spherocity has very unique capability to separate the events based on their geometrical shape, i.e. jetty and isotropic~\cite{Tripathy:2019blo,Khuntia:2018qox}. After its successful implementation in small collision systems, transverse spherocity can be used as a tool for heavy-ion collisions to differentiate the events as well. It might reveal new and unique results from heavy-ion collisions where the production of a QGP medium is already established. In addition, in heavy-ion collisions, after identifying jetty events, one can study jet shapes, medium modification and jet chemistry in a sophisticated manner.

\begin{figure}[ht]
\includegraphics[scale=0.4]{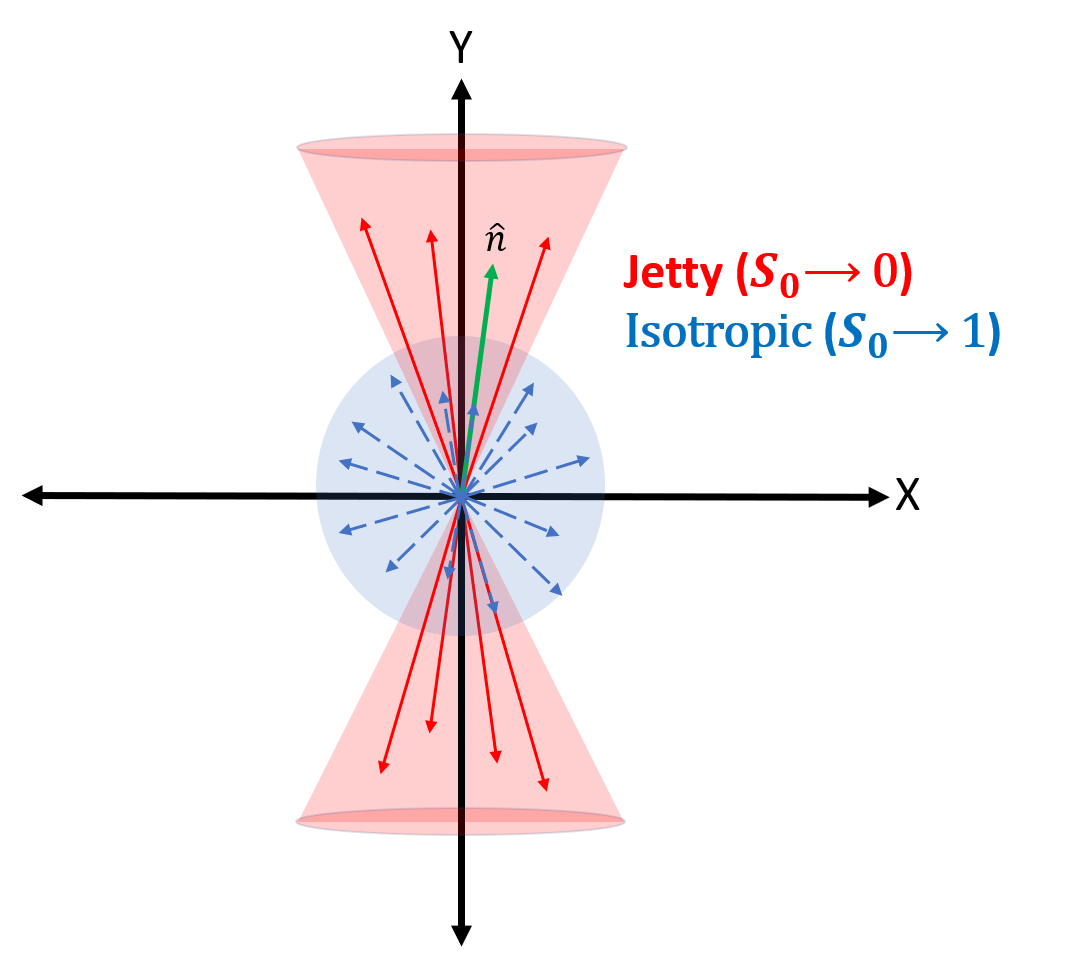}
\caption[]{(Color Online) Schematic picture showing jetty and isotropic events in the transverse plane.}
\label{sp_cart}
\end{figure}

In this work, we use the transverse spherocity for the first time in heavy-ion collisions using A Multi-Phase Transport Model (AMPT). We obtain the transverse momentum spectra, integrated yield, mean transverse momentum for identified particles in Xe-Xe collisions at $\sqrt{s_{\rm{NN}}} = 5.44$~TeV and Pb-Pb collisions at $\sqrt{s_{\rm{NN}}} = 5.02$~TeV. We also study the dependence of elliptic flow on the event types. Elliptic flow is caused by the initial spatial anisotropy in the system produced in any non-central collision and it plays an important role to understand the collective motion and bulk property of the QGP. It is defined as the second-order Fourier component of the particle azimuthal distribution, which provides information about the initial state geometrical anisotropy and the transport properties of created medium in heavy-ion collisions~\cite{v2}. Another important feature of $v_{2}$ is the number of constituent quark (NCQ) scaling, which interprets the dominance of the quark degrees of freedom at early stages of the collision. Recently, from LHC results it seems that $v_{2}$ does not follow the NCQ scaling at LHC energies~\cite{v2_ALICE1, v2_ALICE2} for intermediate or high momentum. It would be very interesting to study these properties of $v_{2}$ in different event shapes. 

The paper is organised as follows. We begin with a brief introduction and motivation for the study in Section~\ref{intro}. In Section~\ref{section2}, the detailed analysis methodology along with brief description about AMPT are given. Section~\ref{section3} discusses about the results and finally they are summarized in Section~\ref{section4}.

\section{Event Generation and Analysis Methodology}
\label{section2}

In this section, we begin with a brief introduction on AMPT model. Then, we proceed to define the elliptic flow and transverse spherocity as an event shape analysis tool.

\subsection{A Multi-Phase Transport (AMPT) Model}
\label{formalism}
A Multi-Phase Transport Model contains four components namely, initialization of collisions, parton transport after initialization, hadronization mechanism and hadron transport~\cite{AMPT2}.  The initialization of the model follows HIJING model~\cite{ampthijing}. The differential cross-section of the produced minijet particles in $pp$ collisions from HIJING is given by,
\begin{eqnarray}
\frac{d\sigma}{dp_{\rm T}^2\,dy_1\,dy_2}=&K&\,\sum_{a,b}x_1f_a(x_1,p_{T1}^2)\,x_2f_2(x_2,p_{T2}^2)\nonumber\\
&\times&\frac{d\hat{\sigma}_{ab}}{d\hat{t}}\,,
\end{eqnarray}
where $\sigma$ is cross-section of produced particles and $\hat{t}$ is the momentum transfer during partonic interactions in $pp$ collisions. $x_i$'s are the momentum fraction of the mother protons which are carried by interacting partons and $f(x, p_{\rm T}^2)$ is the parton density functions (PDF). 
The produced partons calculated in $pp$ collisions is then converted into A-A and $p$-A collisions and they are incorporated via parametrized shadowing function and nuclear overlap function using inbuilt Glauber model within HIJING. Similarly, initial low-momentum partons are produced from parametrized coloured string fragmentation mechanisms. Initial low-momentum partons are separated from high momentum partons by momentum cut-off. The produced particles are initiated into parton transport part, ZPC~\cite{amptzpc}. The transport of the quarks and gluons using Boltzmann transport equation is given by,
\begin{eqnarray}
p^{\mu}\partial_{\mu}f(x,p,t)=C[f].
\end{eqnarray}
Here, $p^{\mu}$, $f(x,p,t)$ and $C[f]$ are four momentum, parton distribution function and collision integral, respectively.
The leading order equation showing interactions among partons is approximately given by,
\begin{equation}
\frac{d\hat{\sigma}_{gg}}{d\hat{t}}\approx \frac{9\pi\alpha_s^2}{2(\hat{t}-\mu^2)^2}\,.
\end{equation}
Here, $\sigma_{gg}$ is the gluon scattering cross-section, $\alpha_s$ is the strong coupling constant used in the above equation, and $\mu^2$ is the cutoff used to avoid infrared divergences which can occur if the momentum transfer, $\hat{t}$, goes to zero during scattering. In the String Melting version of AMPT (AMPT-SM), melting of colored strings into low momentum partons take place at the start of the ZPC. It is calculated using Lund FRITIOF model of HIJING. The resulting partons undergo multiple scatterings which take place when any two partons are within distance of minimum approach. It is given by $\displaystyle d\,\leq\,\sqrt{\sigma/\pi}$, where $\sigma$ is the scattering cross-section of the partons. In AMPT-SM, the transported partons are finally hadronized using coalescence mechanism~\cite{amptreco}.
The coalescence phenomenon takes place using the following equation (for e.g. meson),
\begin{eqnarray}
\frac{d^3N}{d^3p_M}=g_M\int{d^3x_1d^3x_2d^3p_1d^3p_2\,f_q(\vec{x}_1,\vec{p}_1)}f_{\bar{q}}(\vec{x}_2,\vec{p}_2)\nonumber\\
                            \delta^3(\vec{p}_M-\vec{p}_1-\vec{p}_2)\,f_M(\vec{x}_1-\vec{x}_2,\vec{p}_1-\vec{p}_2),
\end{eqnarray}
where, $g_M$ is the meson degeneracy factor, $f_q$'s are the quark distributions after the evolution. $f_M$ is the coalescing function called as Wigner functions~\cite{Greco:2003mm}.

\begin{table*}[ht!]
\begin{center}
\caption{Lowest 20 \% (jetty) and highest 20\% (isotropic) cuts on spherocity distribution. Here, jetty cut specifies 0 to the given value while the isotropic cut begins from the reported value to 1.} 
\label{tab:1}
\begin{tabular}{ |p{2cm}|p{2cm}|p{2cm}|p{2cm}|p{2cm}|p{2cm}|}
 \hline
  &\multicolumn{2}{|c|}{Xe-Xe, $\sqrt{s_{\rm NN}} = 5.44$ TeV} & \multicolumn{2}{|c|}{Pb-Pb, $\sqrt{s_{\rm NN}} = 5.02$ TeV}\\
  \hline
 Centrality (\%) & Jetty & Isotropic & Jetty & Isotropic\\
\hline
0-10		&     --       &    --        & 0.88265   &	0.95445 \\
10-20	&     --       &    --        & 0.81125   &    0.91325  \\
20-30	&0.77385	&0.90365	& 0.75955	  &	0.88355 \\
30-40	&0.74345	&0.88895	& 0.73445	  &	0.86795 \\
40-50	&0.72745	&0.88235	& 0.71585	  &	0.86485 \\
50-60	&0.71845	&0.87905	& 0.70965	  &	0.87035 \\
60-70	&0.70655	&0.87465	& 0.70685	  &	0.87325 \\	
70-100	&0.51535	&0.81455	& 0.53425	  &	0.82325 \\
 \hline
 \end{tabular}
 \end{center}
\end{table*}

The produced hadrons further undergo final evolution in ART mechanism~\cite{amptart1, amptart2} via meson-meson, meson-baryon and baryon-baryon interactions. There is also a default version of AMPT, where instead of coalescing the partons, fragmentation mechanism using Lund fragmentation parameters $a$ and $b$ are used for hadronizing the transported partons. However, the particle flow and spectra at the mid-$p_{\rm T}$ regions are well explained by quark coalescence mechanism for hadronization~\cite{ampthadron1,ampthadron2,ampthadron3}. We have used AMPT-SM mode for all of our calculations and we have used the AMPT version 2.26t7 (released: 28/10/2016) in our current work. The AMPT settings in the current work, are exactly the same as reported in Ref.~\cite{Tripathy:2018bib}. For the input of impact parameter values for different centralities in Xe-Xe and Pb-Pb collisions, we have used Ref.~\cite{Loizides:2017ack}. One should note here that, high centrality collisions corresponds to low impact parameter values and higher final state charged-particle multiplicity ($\langle dN_{\rm ch}/d\eta \rangle$). Although the concept of centrality is widely used in heavy-ion collisions, in view of a final state multiplicity scaling across collisions species, that is observed at the LHC energies, we may use centrality and $\langle dN_{\rm ch}/d\eta \rangle$ variably in this work.

\subsection{Elliptic Flow}

The anisotropic flow of different order can be characterized by the coefficients ($v_n$), which are obtained from a Fourier expansion of the momentum distribution of the charged particles. It is given by,
\begin{eqnarray}
E\frac{d^3N}{d^3p}=\frac{d^2N}{2\pi p_{\rm T}dp_{\rm T}dy}\bigg(1+2\sum_{n=1}^\infty v_n \cos[n(\phi -\psi_n)]\bigg)\,.\nonumber\\
\label{eq5}
\end{eqnarray}
Here, $\phi$ is the azimuthal angle in the transverse momentum plane and $\psi_n$ is the n$^{\text{th}}$ harmonic event plane angle~\cite{v2eventplane}. In the current work, elliptic flow is calculated with respect to the reaction plane by taking $\psi_{n}$ = 0. This implies event plane coincides with the reaction plane. Although it is non-trivial in experiments but AMPT provides the freedom to exactly define the event plane for a collision. Taking $n$ = 2 in Eq.~\ref{eq5} gives the second order harmonics in the expansion and its coefficient, $v_2$ is calculated to provide the measure of the elliptic flow or azimuthal anisotropy. Thus, $v_{2}$ is defined as:
 \begin{eqnarray}
 v_{2} = \langle \cos(2\phi)\rangle
\end{eqnarray}
Currently, we use Eq.~\ref{eq5} with $\psi_{n}$ = 0 to calculate the elliptic flow. However, to compare with experimental data, we are now moving to two-particle correlation method to calculate the elliptic flow. The two-particle correlation method has an added advantage as by construction, it would remove the non-flow effects in the elliptic flow.
\subsection{Transverse Spherocity}
\begin{figure}[ht]
\includegraphics[scale=0.42]{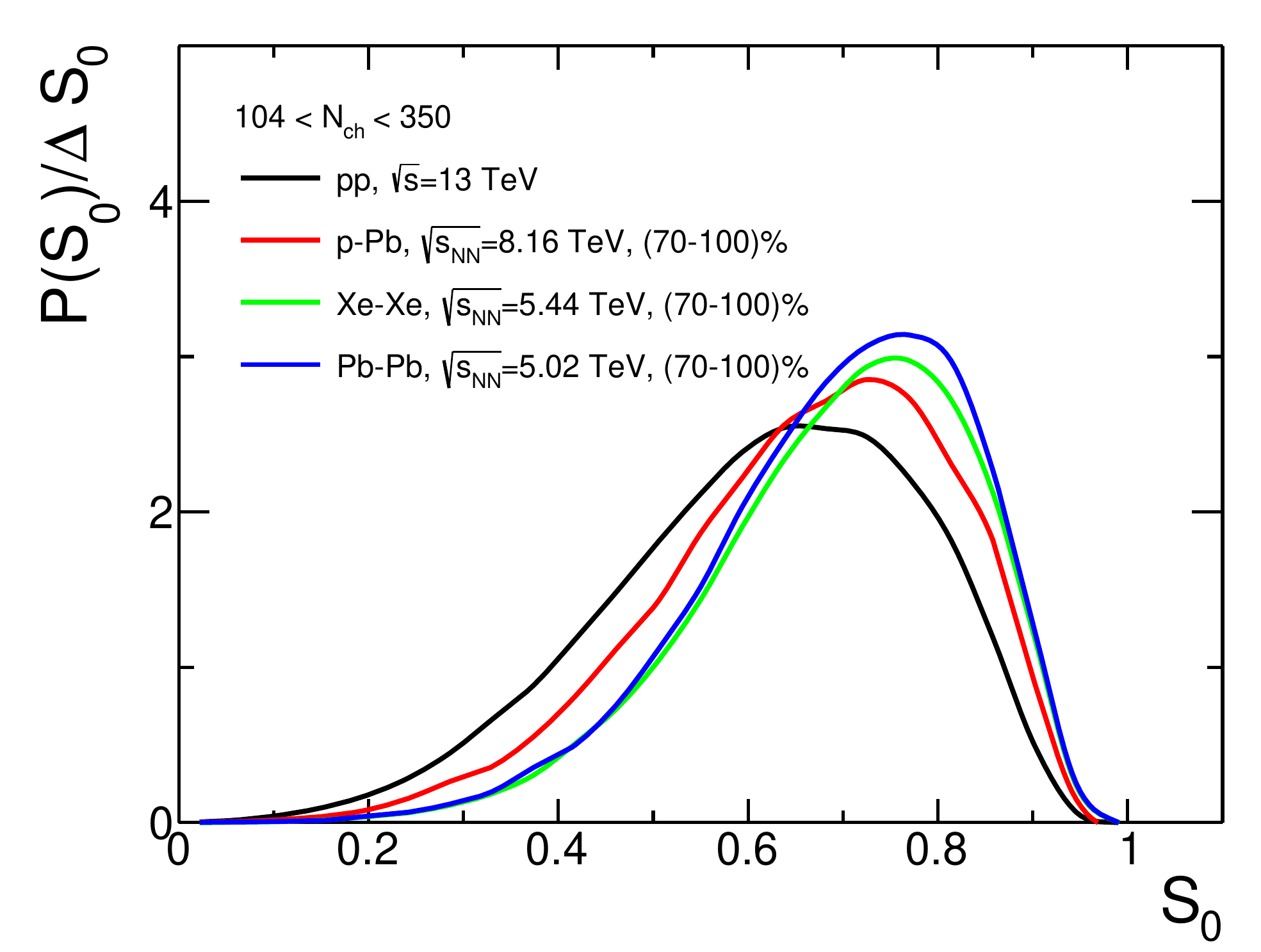}
\caption[]{(Color Online) Spherocity distributions for $pp$, $p$-Pb, Xe-Xe and Pb-Pb collisions in common final state charged-particle multiplicity of 104 $< N_{ch} < 350$.}
\label{sp_common}
\end{figure}
Transverse spherocity is an event property which is defined for a unit vector $\hat{n} (n_{T},0)$ that minimizes the ratio~\cite{Cuautle:2014yda, Cuautle:2015kra}:
\begin{eqnarray}
S_{0} = \frac{\pi^{2}}{4} \bigg(\frac{\Sigma_{i}~\vec p_{T_{i}}\times\hat{n}}{\Sigma_{i}~p_{T_{i}}}\bigg)^{2}.
\label{eq7}
\end{eqnarray}

\begin{figure*}[ht!]
\begin{center}
\includegraphics[scale=0.29]{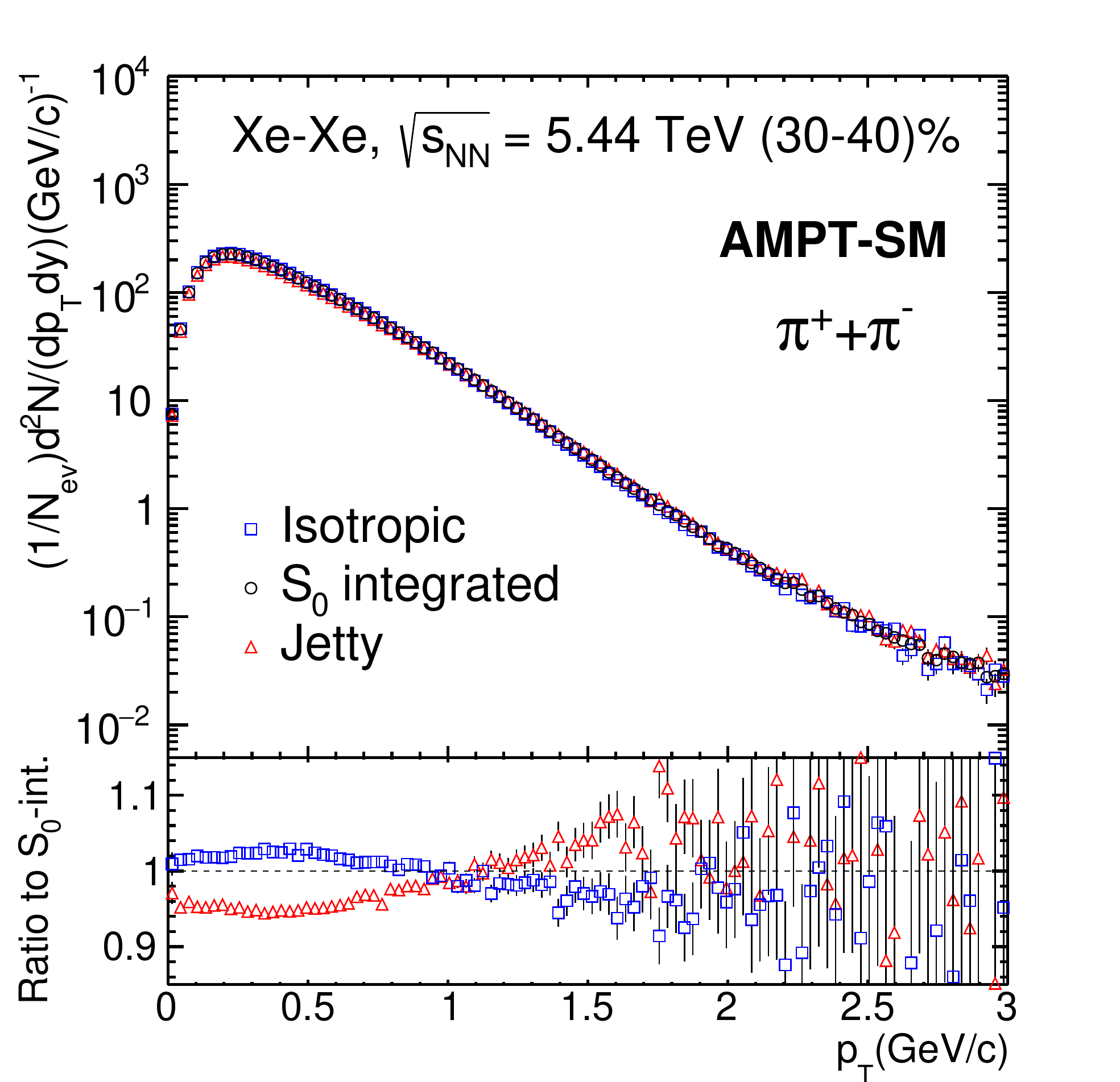}
\includegraphics[scale=0.29]{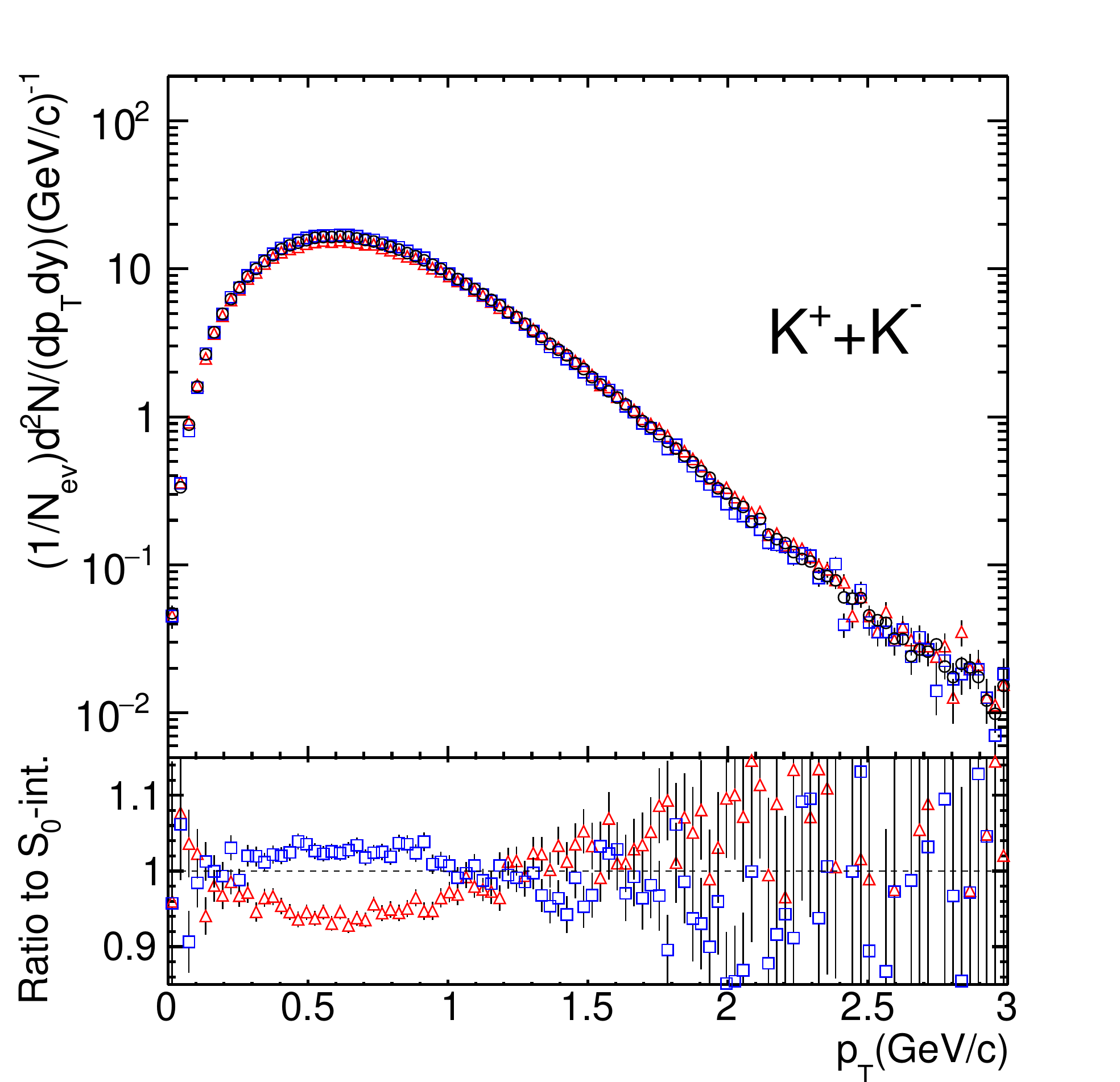}
\includegraphics[scale=0.29]{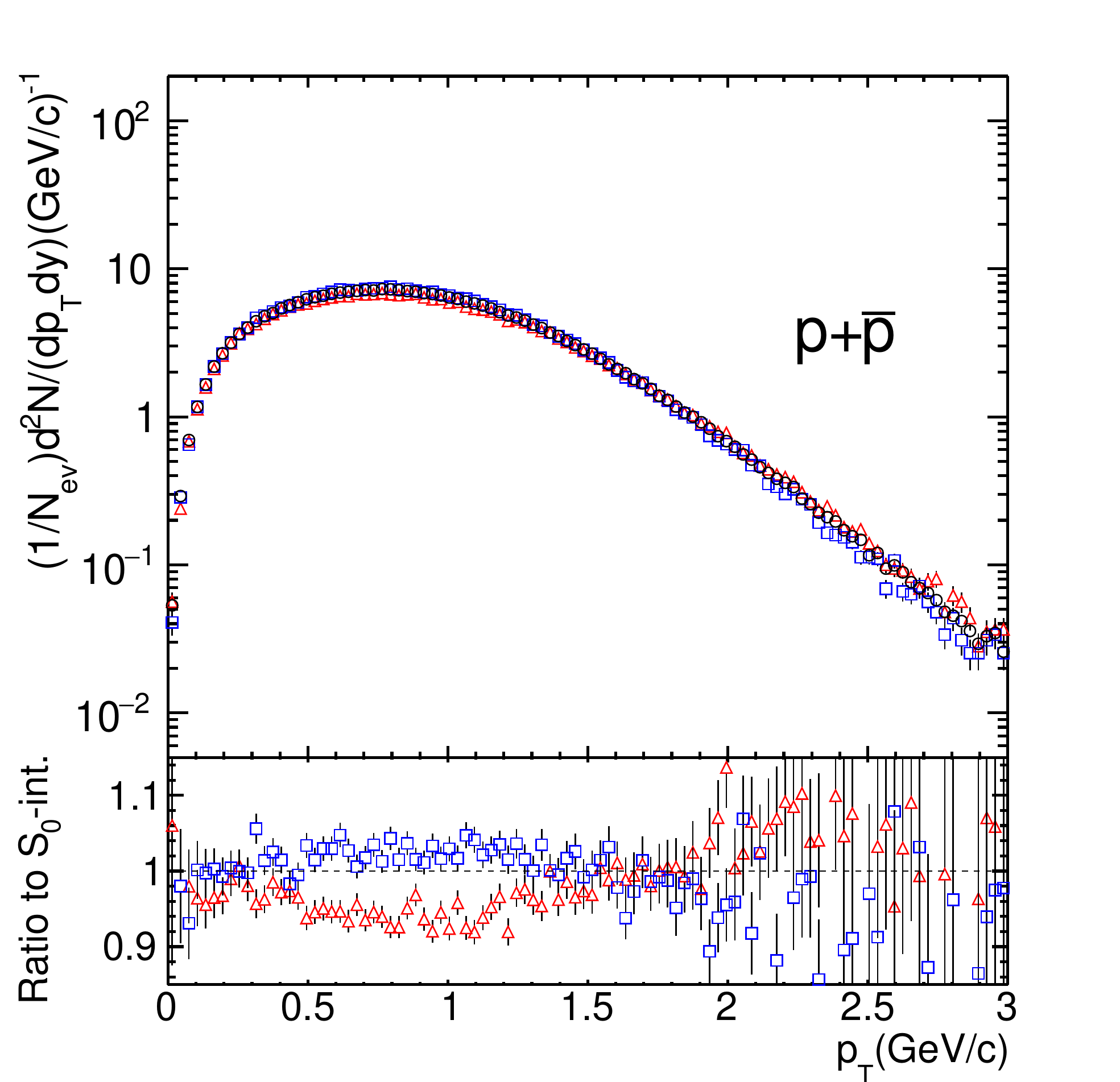}
\caption[width=18cm]{(Color Online) Top plot: $p_{\rm T}$-spectra for pions, kaons and protons in Xe-Xe collisions at (30-40)\% centrality with isotropic, $S_0$-integrated and jetty events. Bottom plot: Ratio of $p_{\rm T}$-spectra for isotropic and jetty events to the $S_0$-integrated events. }
\label{pTSpectra1}
\end{center}
\end{figure*}
\begin{figure*}[ht!]
\begin{center}
\includegraphics[scale=0.29]{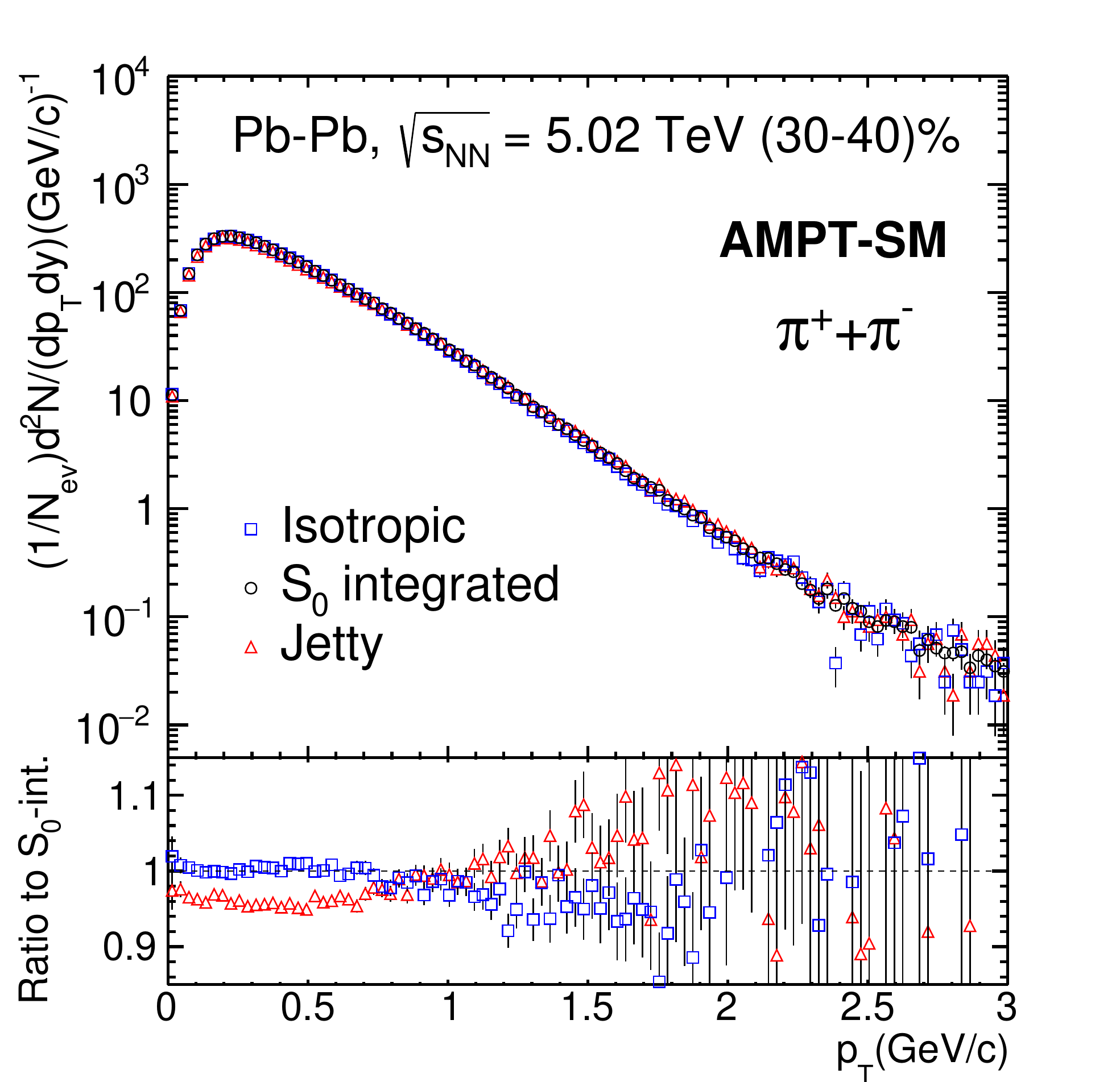}
\includegraphics[scale=0.29]{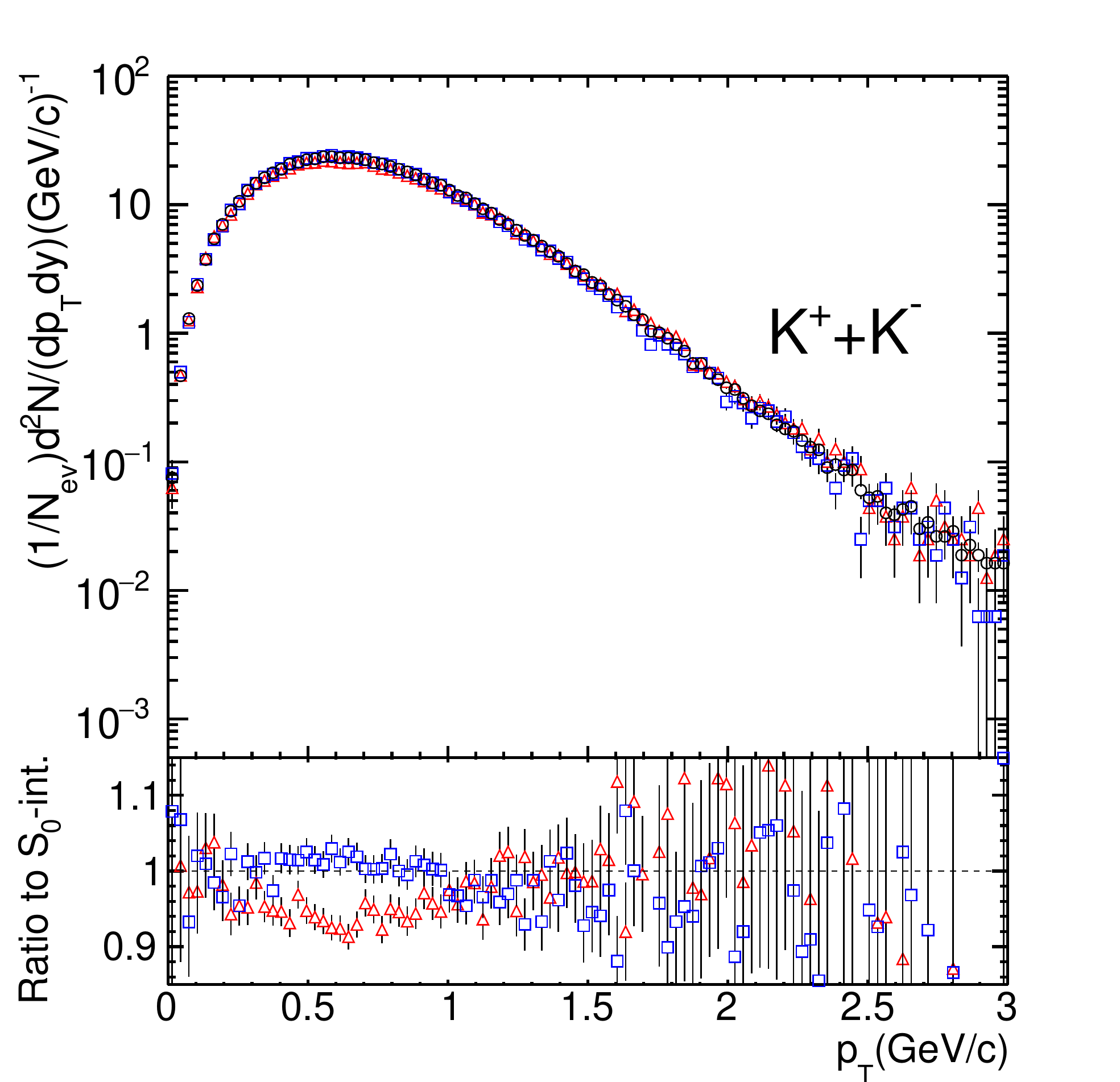}
\includegraphics[scale=0.29]{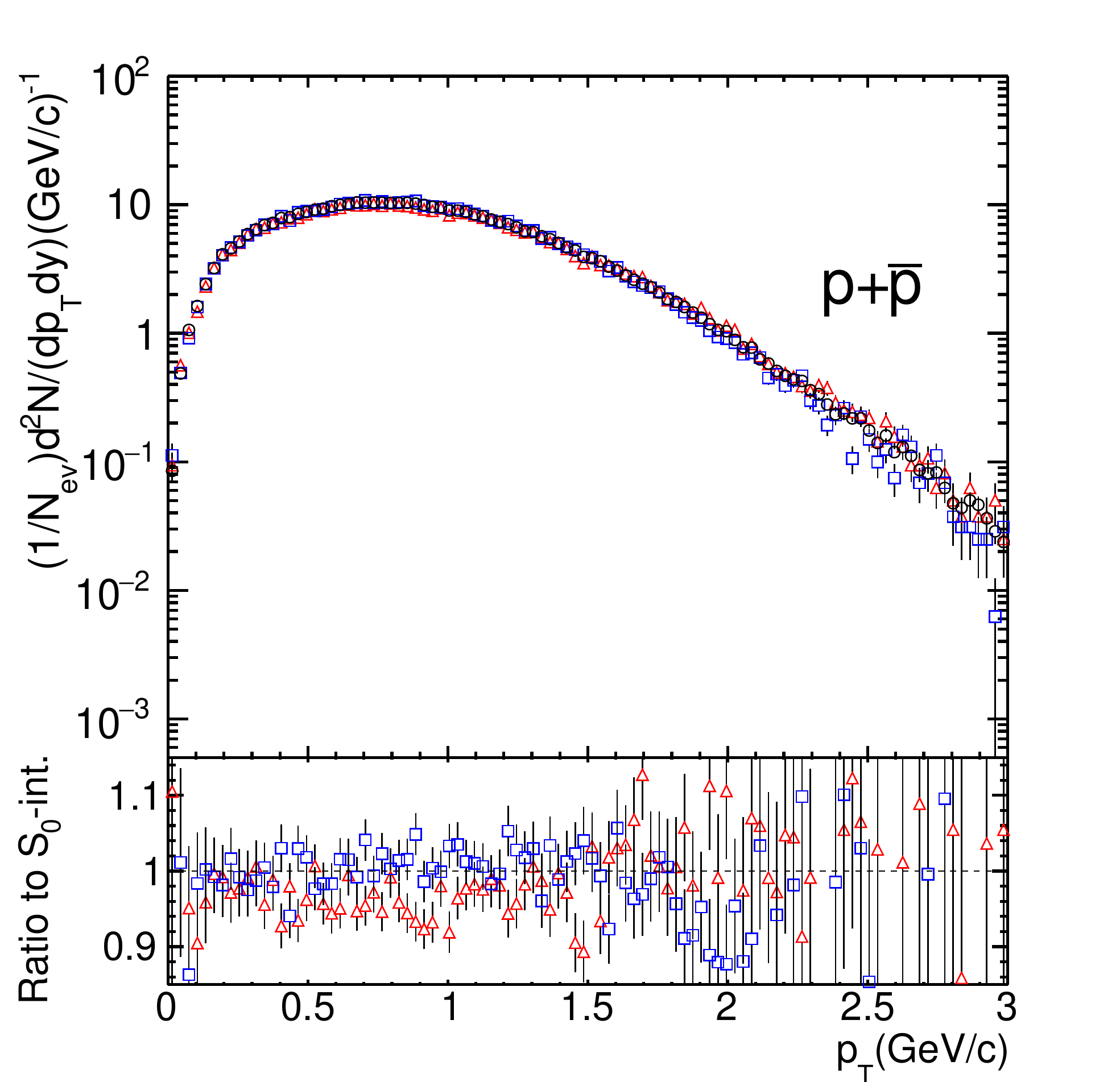}
\caption[width=18cm]{(Color Online) Top plot: $p_{\rm T}$-spectra for pions, kaons and protons in Pb-Pb collision at (30-40)\% centrality with isotropic, $S_0$-integrated and jetty events. Bottom plot: Ratio of $p_{\rm T}$-spectra for isotropic and jetty events to the $S_0$-integrated events. }
\label{pTSpectra2}
\end{center}
\end{figure*}

By restricting it to transverse plane, transverse spherocity becomes infrared and collinear safe~\cite{Salam:2009jx}. By construction, the extreme limits of transverse spherocity are related to specific configurations of events in transverse plane. The value of transverse spherocity ranges from 0 to 1, which is ensured by multiplying the normalization constant $\pi{^2}/4$ in Eq.~\ref{eq7}. Transverse spherocity becoming 0 means, the events are pencil-like (back-to-back structure) and called as jetty events while 1 would mean the events are isotropic as shown in Fig.~\ref{sp_cart}. The jetty events are usually the hard events while the isotropic events are the result of soft processes.

\begin{figure*}[ht!]
\centering
\includegraphics[width=16cm, height=5cm]{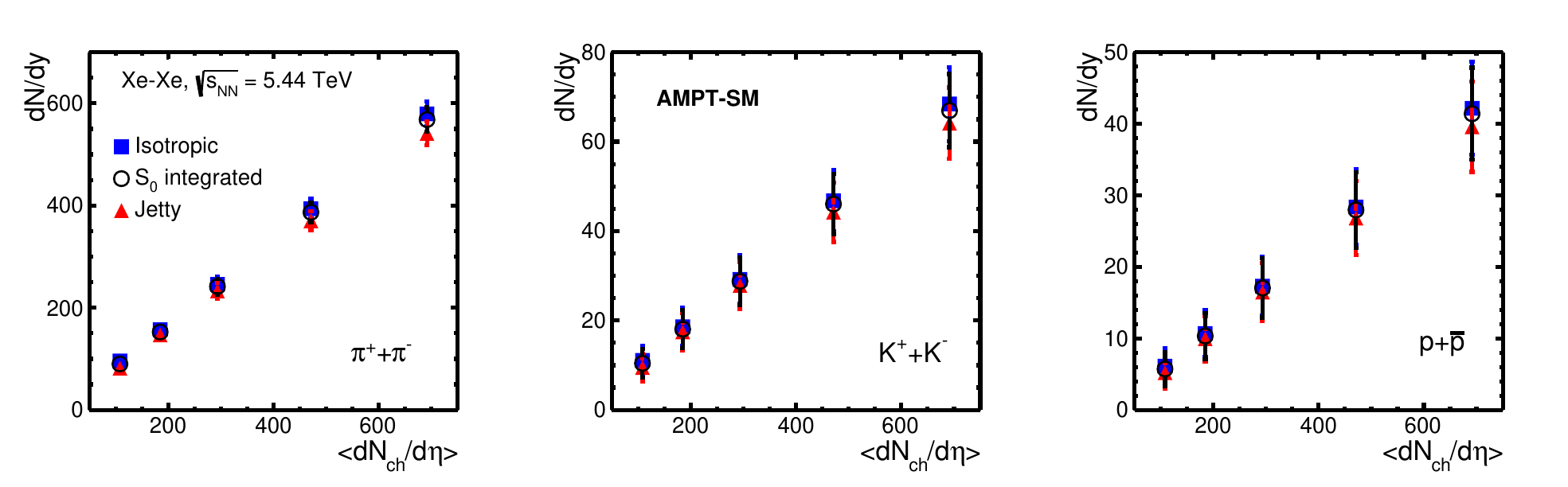}
\includegraphics[width=16cm, height=5cm]{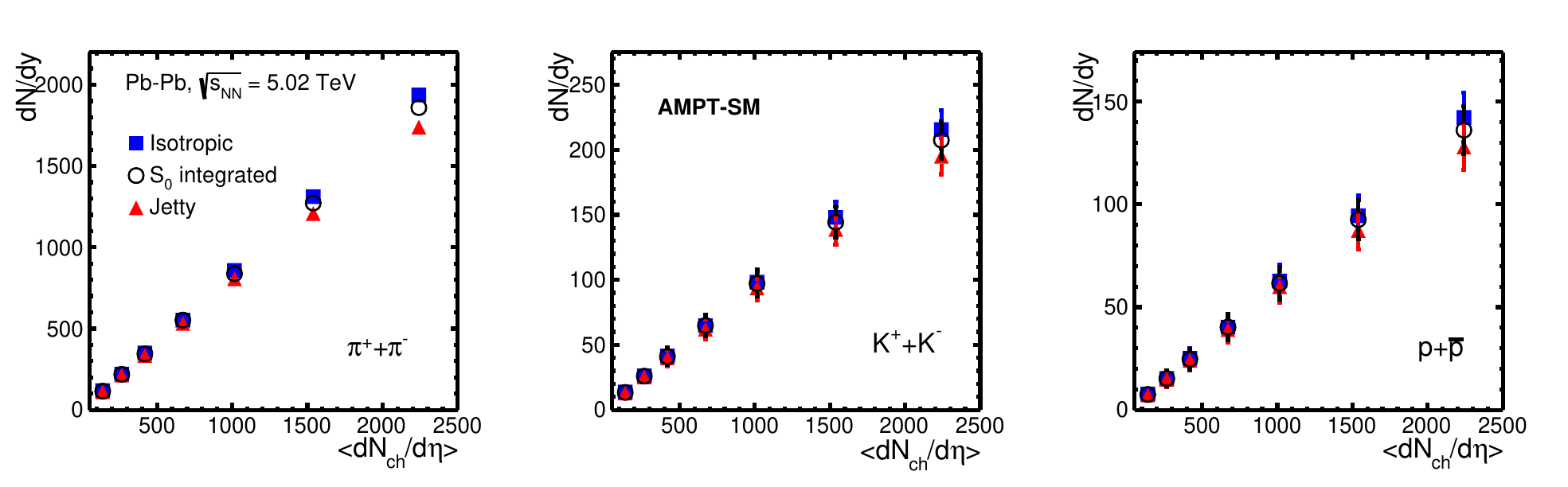}
\caption[]{(Color Online) Multiplicity dependence of integrated yield of pions, kaons and protons for isotropic, $S_0$-integrated and jetty events in Xe-Xe and Pb-Pb collisions using AMPT-SM event generator.}
\label{IntYield}
\end{figure*}

Here onwards, for the sake of simplicity the transverse spherocity is referred as spherocity. In our analysis, we have calculated the spherocity values corresponding to different events for a given system and energy. To disentangle the jetty and isotropic events from the average-shaped events, we have applied spherocity cuts on our generated events. The spherocity distributions are selected in the pseudo-rapidity range of $|\eta|<0.8$ with a minimum constraint of 5 charged particles with $p_{\rm{T}}>$~0.15~GeV/$c$ to recreate the similar conditions as in ALICE experiment at the LHC. The jetty events are those events having spherocity values in the lowest 20 percent and the isotropic events are those occupying the highest 20 percent in the spherocity distribution of the all events. The spherocity distributions for $pp$ collisions at $\sqrt{s} = 13$ TeV, $p$-Pb collisions at $\sqrt{s_{\rm NN}}$ = 8.16 TeV, Xe-Xe collisions at $\sqrt{s_{\rm NN}} = 5.44$ TeV and Pb-Pb collisions at $\sqrt{s_{\rm NN}} = 5.02$ TeV with a common final state charged particle multiplicity are shown in Fig.~\ref{sp_common} as a demonstration plot. We observe that, even at common charged-particle multiplicity, the spherocity distributions are shifted towards more isotropic events with increasing system size. This indicates that apart from final state charged-particle multiplicity, the collision system and collision species still have roles to play on the event types. This behavior was also seen in several phenomenological analysis of experimental data~\cite{Sahu:2020nbu,Sahu:2019tch}. 

We now proceed for the estimation of transverse momentum spectra, integrated yield, mean transverse momentum and azimuthal anisotropy in different spherocity classes at the mid-rapidity ($|\eta|<$  0.8) for Xe-Xe and Pb-Pb collisions at the LHC energies from AMPT. 
\begin{figure*}[ht!]
\centering
\includegraphics[width=18cm, height=6cm]{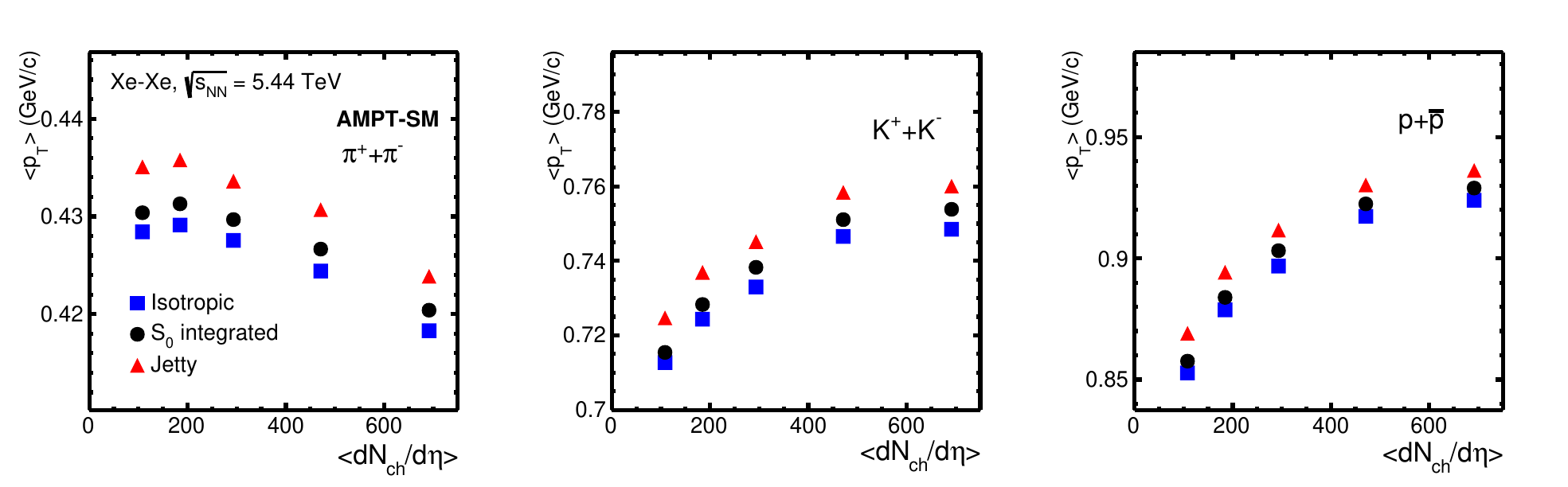}
\includegraphics[width=18cm, height=6cm]{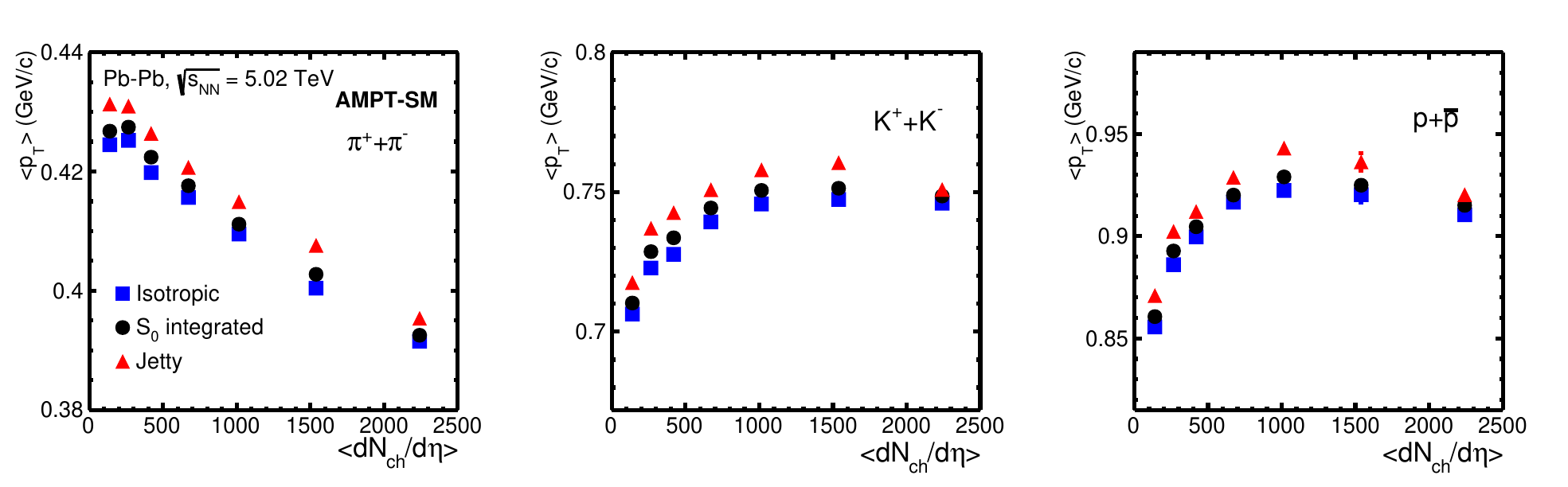}
\caption[]{(Color Online) Multiplicity dependence of mean transverse momentum ($\langle p_{\rm T} \rangle$) of pions, kaons and protons for isotropic, $S_0$-integrated and jetty events in Xe-Xe and Pb-Pb collisions using AMPT-SM model.}
\label{meanpT}
\end{figure*}

\section{Results and Discussions}
\label{section3}
\subsection{$p_{\rm T}$-Spectra}
Figures \ref{pTSpectra1} and \ref{pTSpectra2} show the $p_{\rm T}$-spectra of pions, kaons and protons for Xe-Xe and Pb-Pb collisions at $\sqrt{s_{\rm NN}}$~=~5.44 and 5.02 TeV for different spherocity classes in (30-40)\% centrality, respectively. The bottom plots show the ratio of $p_{\rm T}$-spectra of isotropic and jetty events to spherocity integrated events. (30-40)\% centrality is chosen as a representative spectra. In low-$p_{\rm T}$ region ($0<p_{\rm T}<1.5$), the number of pions produced is much higher than the heavier particles i.e. kaons and protons, suggesting higher production cross-section for lower mass particles. We have implemented the spherocity analysis to separate and distinguish isotropic and jetty events. We observe that, in low-$p_{\rm T}$ region, particle production is dominated by isotropic events. But as we move on to slightly higher $p_{\rm T}$, we see that, there is a certain $p_{\rm T}$ value at which the $p_{\rm T}$-spectra of both isotropic and jetty events cross each other. We call it as the ``crossing point". When we move further to higher $p_{\rm T}$ scale, there is a switch-over between the two types of events and particle production is now mostly due to jetty events while isotropic events contribute less afterwards.
The interesting point to notice is that, the crossing point is mass dependent i.e. it shifts towards higher  $p_{\rm T}$ as the mass of the particle increases. We see here, as $m_\pi<m_K<m_p$, the crossing point shifts towards higher $p_{\rm T}$ as we move from pion to proton and kaon is intermediate. This is behavior is an indication of possible collectivity in heavy-ion collisions. High-$p_{\rm T}$ region is pQCD dominated and the shift of crossing point towards high-$p_{\rm T}$ for high mass particles is an indication that massive particles are produced through pQCD processes.

\subsection{Mean Transverse Momenta ($\langle p_{\rm T} \rangle$)}
In Fig.~\ref{meanpT}, the $\langle p_{\rm T} \rangle$ as a function of charged-particle multiplicity for pions, kaons and protons at mid-rapidity in Xe-Xe and Pb-Pb collisions for isotropic, $S_0$-integrated and jetty events have been shown. We observe that, $\langle p_{\rm T} \rangle$ is clearly dependent on the final state charged-particle multiplicity. In case of pions, $\langle p_{\rm T} \rangle$ keeps on increasing as we move from central to peripheral collisions and saturates for peripheral collisions. But, in case of kaons and protons, we observe that, $\langle p_{\rm T} \rangle$ has higher value in central collisions and keeps on decreasing as we move towards peripheral collisions. The different behavior of $\langle p_{\rm T} \rangle$ for pions compared to other particles could be due to the contribution of resonance decays to production of pions, which is higher for central collisions compared to peripheral collisions. Similar trend is observed for both Xe-Xe and Pb-Pb collisions.

Now, as we have implemented event separation using spherocity as our tool, we can now understand which type of events carry more $p_{\rm T}$ in the system. Jetty events are usually are the results of hard-QCD processes and it results in the production of less number of particles with high $p_{\rm T}$. In contrast, isotropic events are because of soft-QCD processes and they yield more number of particles. As the momentum of the system should remain conserved and the sum of momenta of final state particles should always be equal to initial momentum, therefore, isotropic events should carry less $p_{\rm T}$ as compared to jetty events. As integrated yield is found to be higher for isotropic compared to jetty events, $\langle p_{\rm T} \rangle$ being higher for jetty events is in accordance with energy-momentum conservation. This has been observed in our work. As we can see, for both Xe-Xe and Pb-Pb collisions at different centralities, isotropic events have less $\langle p_{\rm T} \rangle$ than jetty i.e. jetty events give more momentum to the outgoing particles.  However, while going from central to peripheral collisions, the $\langle p_{\rm T} \rangle$ for all the event types seems similar.

\subsection{Integrated Yields (dN/dy)}
Figure~\ref{IntYield} shows that the integrated yield of identified particles for Xe-Xe and Pb-Pb collisions in different spherocity classes. As expected, the integrated yield decreases from central to peripheral collisions. It is observed that the integrated yield highly depends on the spherocity classes for most central heavy-ion collisions and the dependence decreases while going towards peripheral collisions. The contribution from isotropic events to the integrated yield is higher than that from jetty events. This is understood by the fact that the integrated yield is dominated by the low-$p_{\rm T}$ particles and the particle production at low-$p_{\rm T}$ is dominated by particle produced from isotropic events. This behavior is also evident in Figs.~\ref{pTSpectra1} and~\ref{pTSpectra2}.

\begin{figure*}[ht!]
\centering
\includegraphics[width=18cm, height=6cm]{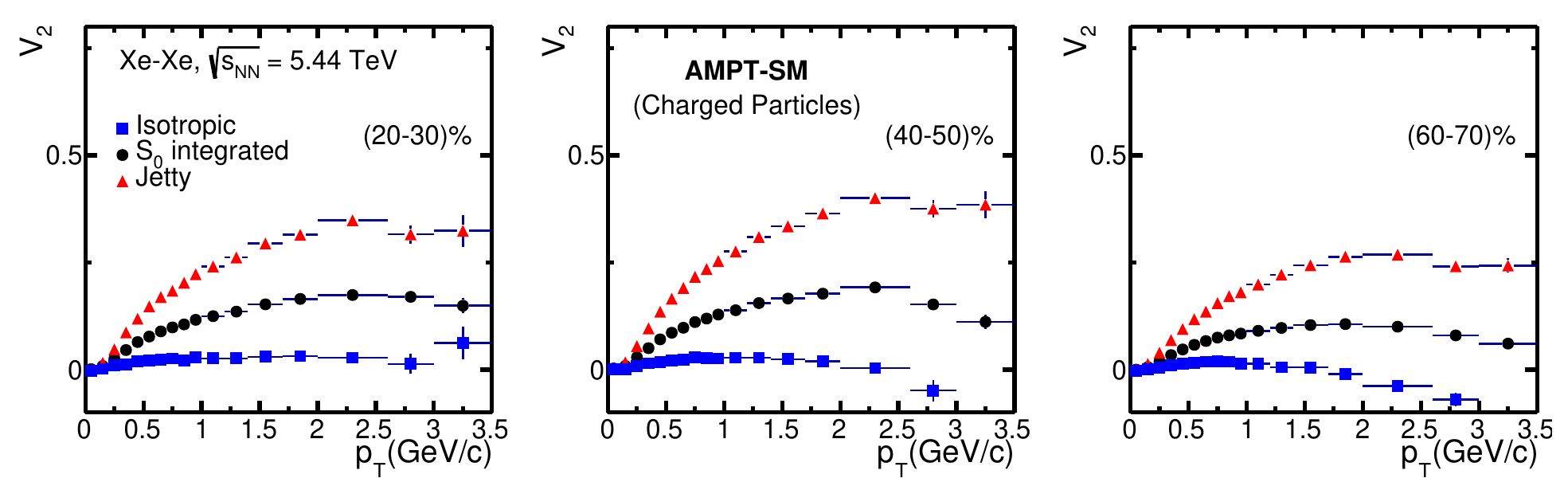}
\includegraphics[width=18cm, height=6cm]{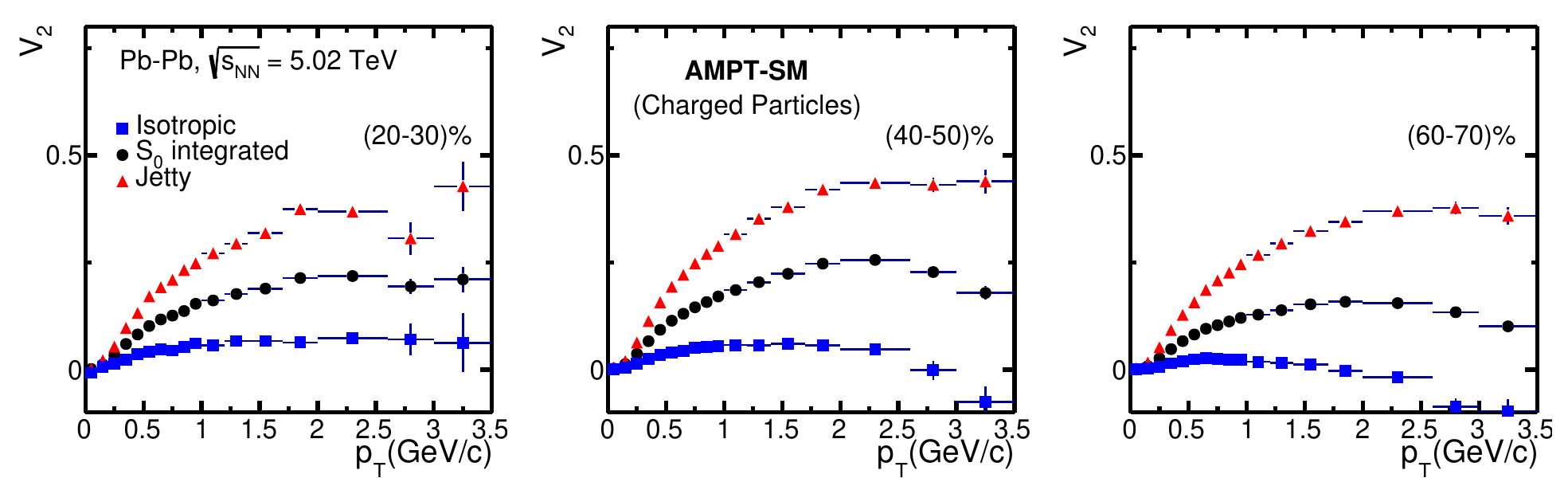}
\caption[width=18cm]{(Color Online) Centrality dependence of elliptic flow for Xe-Xe and Pb-Pb collisions at different centralities using AMPT-SM model. Event shape dependencies are shown for various centralities.}
\label{v2_centrality1}
\end{figure*}

\subsection{Elliptic Flow}
\subsubsection{Centrality Dependence}
\label{Centrality_Dependence}

\begin{figure}
\begin{subfigure}
\centering
\includegraphics[width=8cm,height=6cm]{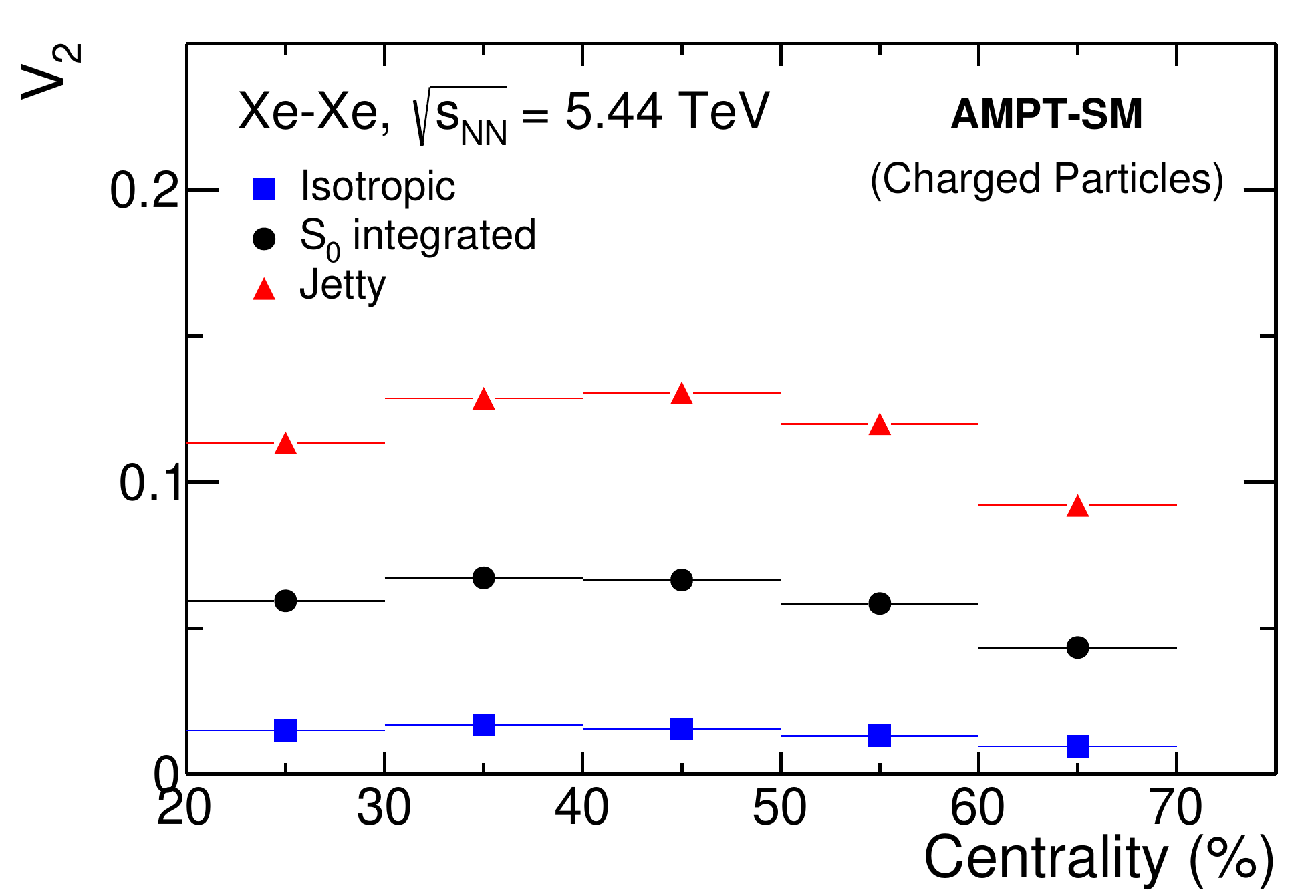}
\end{subfigure}
\hfill
\begin{subfigure}
\centering
\includegraphics[width=8 cm,height=6cm]{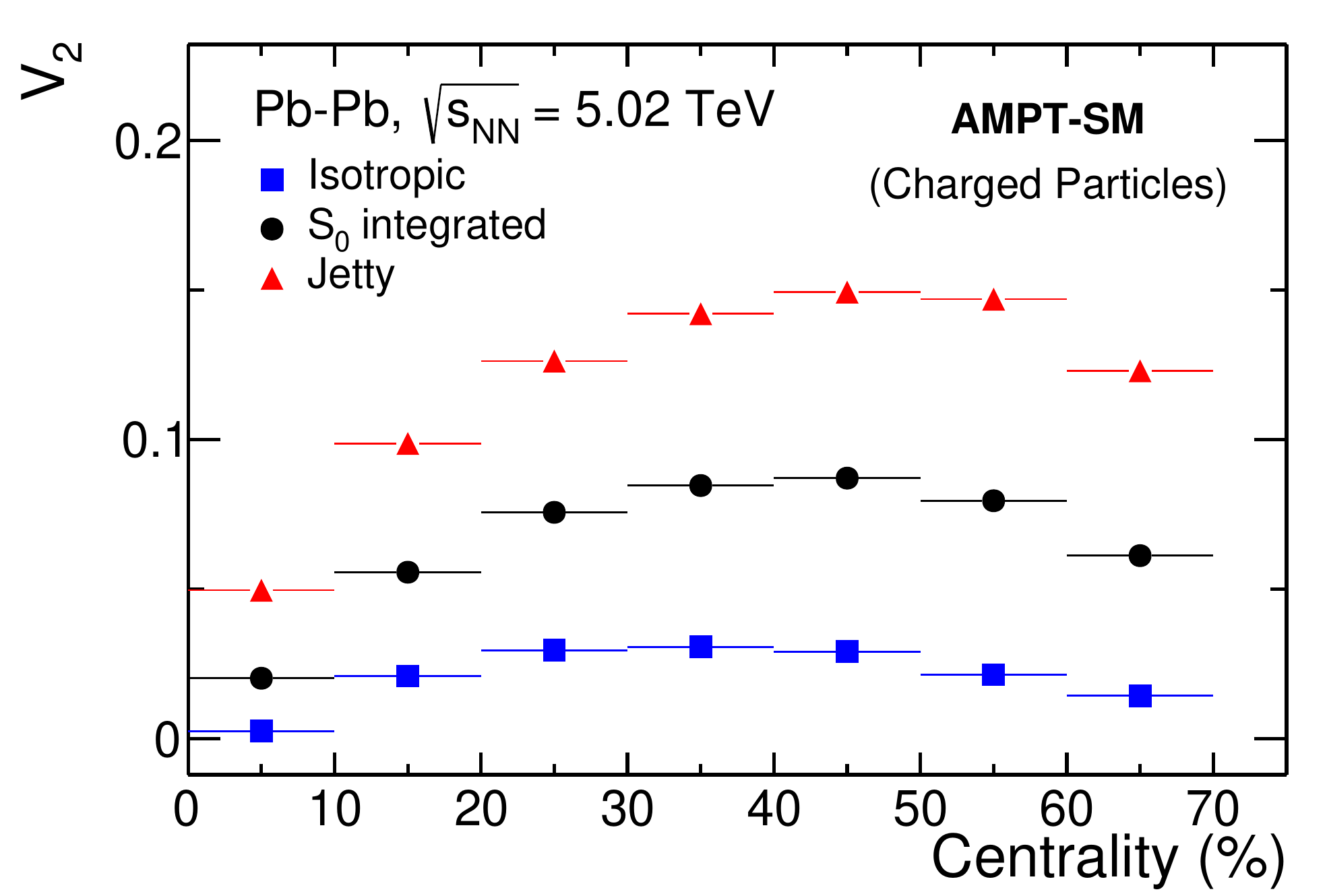}
\end{subfigure}
\caption[]{(Color Online) Centrality dependence of elliptic flow for Xe-Xe and Pb-Pb collisions at different centralities using AMPT-SM model.}
\label{v2_centrality2}
\end{figure}

We have estimated the elliptic flow ($v_2$) for all charged particles in mid-rapidity for different charged-particle multiplicities in Xe-Xe and Pb-Pb collisions for isotropic, $S_0$-integrated and jetty events, which is shown in Fig.~\ref{v2_centrality1}. $v_2$ is an initial state property and gives information about the momentum space anisotropy or azimuthal anisotropy in the medium at the earliest time of its formation. In both Xe-Xe and Pb-Pb collisions, we observe finite $v_2$. We also see that, $v_2$ is strongly dependent on the final state charged-particle multiplicities. As we move from central to peripheral collisions, $v_2$ keeps on increasing and becomes maximum for mid-central collisions. However, its value decreases if we move further towards peripheral collisions. In Fig.~\ref{v2_centrality2}, it is quite evident. For most central collisions, the system has less spatial anisotropy, thus it has less value of $v_2$. In most peripheral collisions, the nuclear overlap region at the collision point decreases and the size as well as the density of participating partons also decrease. So, less number of particles emerge from this type of collisions can not carry the effect of $v_2$ till the final state. But, in case of mid-central collisions, the system has finite spatial anisotropy and the nuclear overlap region has enough participants, hence the produced system has the maximum $v_2$, which can be seen in the Fig. \ref{v2_centrality1}. 

Now, coming to the event types, we observe that, the contribution towards $v_2$ is mostly dominated by jetty events. It can be seen in both Figs. \ref{v2_centrality1} and  \ref{v2_centrality2}. An interesting point to notice is that, isotropic events have almost zero $v_2$. This clearly shows that the types of events have important role towards the initial state anisotropy in the system. Isotropic events showing less $v_2$ is a testimony of transverse spherocity successfully separating jetty and isotropic events through proper event topological selections. The process of isotropization, resulting in events with higher probability of isotropic events diminishes the azimuthal anisotropy in the final state. 

\subsubsection{Collision System Dependence}

\begin{figure*}[ht!]
\includegraphics[width=18cm,height=6cm]{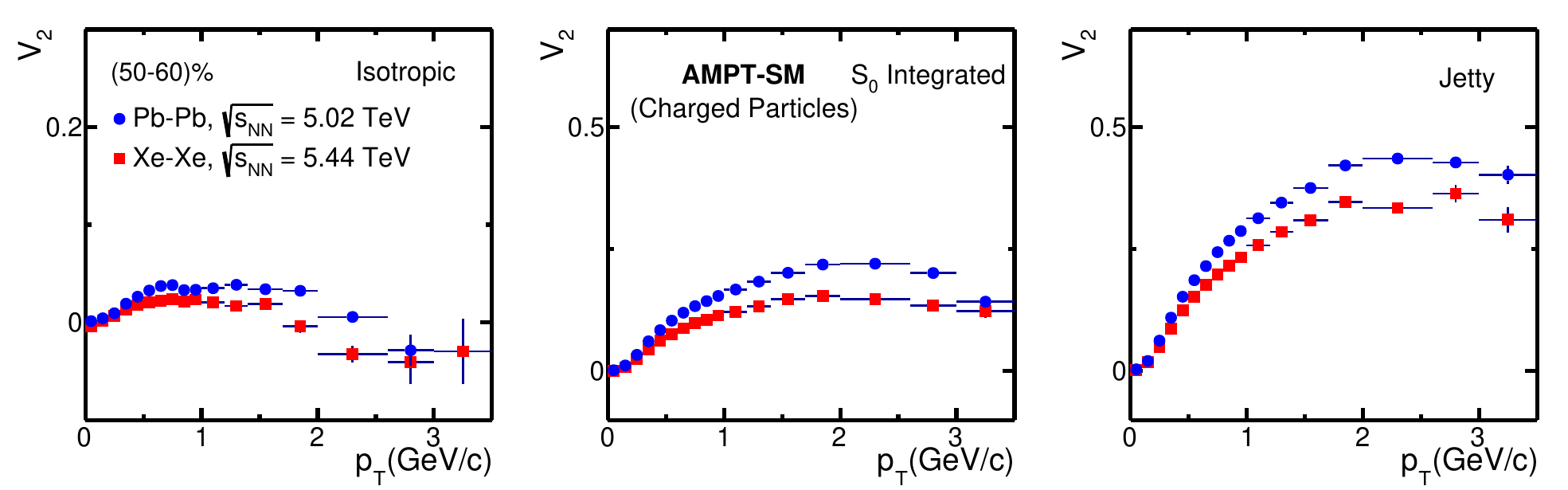}
\caption[]{(Color Online) Collision system dependence of elliptic flow for Xe-Xe and Pb-Pb collisions using AMPT-SM model at the LHC energies.}
\label{system_dependence}
\end{figure*}
In Fig. \ref{system_dependence}, elliptic flow versus transverse momentum ($v_2$ versus $p_{\rm T}$) has been plotted for all charged particles in mid-rapidity in Xe-Xe and Pb-Pb collisions for (50-60)\% class. We observe that azimuthal anisotropy for Pb-Pb system is higher than that of Xe-Xe system. It means, azimuthal anisotropy ($v_2$) is system size dependent. Xe-Xe is a smaller collision system as compared to Pb-Pb, hence it has less elliptic flow. The behavior for spherocity-integrated case qualitatively and quantitatively agrees with Ref.~\cite{Tripathy:2018bib}. The collision energy dependence of both isotropic and jetty events are similar as observed for spherocity-integrated case. However, for isotropic events the difference is not as significant as for other types of events but the Pb-Pb collisions has higher elliptic flow than that of Xe-Xe collisions even for isotopic events. This would indicate that with increasing system size, the isotropic events start contributing towards elliptic flow.

%

\section{Summary and Conclusion}
\label{section4}
In summary,

\begin{enumerate}
\item We report the first implementation of transverse spherocity analysis for heavy-ion collisions at the Large Hadron Collider energies using A Multi-Phase Transport Model (AMPT).\\

\item The results show that transverse spherocity successfully differentiates the heavy-ion collisions event topology based on their geometrical shape i.e. isotropic and jetty.\\

\item At a common charged-particle multiplicity, the spherocity distributions are shifted towards more isotropic events with increasing system size. This indicates that along with final state charged-particle multiplicity, the collision system and collision species have roles to play on the event types. 

\item The indication of collectivity in heavy-ion collisions can be clearly seen while comparing the transverse momentum spectra from jetty and isotropic events.\\

\item The elliptic flow as a function of transverse spherocity shows that the isotropic events have nearly zero elliptic flow. \\

\end{enumerate}

We believe that the results are very encouraging and an experimental exploration in this direction would be highly helpful to understand event topology dependence of system dynamics. All the results presented here will act as very nice baseline for future experimental work. Removing the non-flow contribution from elliptic flow and the number of quark participant scaling for elliptic flow could be the next step forward in these studies. 

\section*{Acknowledgements}
The authors acknowledge the financial supports  from  ALICE  Project  No. SR/MF/PS-01/2014-IITI(G) of Department  of  Science  \&  Technology,  Government of India. R. S. acknowledges the financial supports from DAE-BRNS Project No. 58/14/29/2019-BRNS. The authors would like to acknowledge the usage of resources  of the LHC grid computing facility at VECC, Kolkata.


\end{document}